\documentclass[floatfix,pra,twocolumn,aps,superscriptaddress,showpacs]{revtex4-1}
\usepackage{hyperref}
\hypersetup{
	colorlinks=true,
	citecolor=blue,
	linkcolor=blue,
	urlcolor=blue}
\usepackage{graphicx}
\usepackage[normalem]{ulem}
\usepackage{amsmath}
\usepackage{amsfonts}
\usepackage{amssymb}
\usepackage{epsfig}
\usepackage{subfigure}
\usepackage{mathtools}
\usepackage{braket}
\usepackage[usenames,dvipsnames]{color}
\usepackage{setspace}
\usepackage{bm}
\usepackage{times}
\usepackage{ulem}
\usepackage{dsfont}
\usepackage{xcolor}

\begin{document}
\title{Spatially multiplexed concentric discrete optical vortices: Complex topological structures and unconventional rotational dynamics}
\author{Aditya Narayana Jena} 
\affiliation{Laser Research Lab, Department of Physics, Indian Institute of Technology Ropar, Rupnagar 140001, Punjab, India}
\author{Vishwa Pal}\email{vishwa.pal@iitrpr.ac.in}
\affiliation{Laser Research Lab, Department of Physics, Indian Institute of Technology Ropar, Rupnagar 140001, Punjab, India}

\begin{abstract}
Precise control over the rotational dynamics of structured lights has become a defining objective in contemporary photonics. It plays a central role in governing the functional distribution of optical energy. 
%These rotational dynamics originate from the interplay between spin angular momentum (SAM) and orbital angular momentum (OAM), which together govern energy flow in structured light fields. 
Particularly, orbital angular momentum driven intensity rotation and azimuthal energy flow in vortex beams have emerged as crucial degrees of freedom in light–matter interactions. Leveraging this foundation, we establish a discrete optical vortex (DOV) platform comprising concentric rings of phase-locked lasers that enable precise control over vortex-beam rotation. By engineering the spatial distribution of topological charges (TCs) across the concentric rings of DOVs, we realize shape-invariant asymmetric vortex beams with controllable reversal of intensity rotation, without altering the sign of constituent TCs. This result establishes a new framework for controlling energy flow in vortex beams, beyond the conventional paradigm that links vortex rotation solely to the sign of the TC. Such controllable rotational dynamics opens new avenues for programmable beam steering, advanced optical micromanipulation, information multiplexing, and adaptive structured-light systems.
%This study further reveals that increasing the TC of the outer ring progressively suppresses the angular excursion of peripheral vortices in concentric DOVs, reflecting enhanced confinement and stronger phase averaging. This trend mirrors the reduced angular velocity observed in magnetic skyrmion bags with higher chiral-kink (TC) density, underscoring a broader correspondence between peripheral structural complexity and rotational dynamics.
\end{abstract}
\maketitle
%%%%%%%%%%%%%%%%%%%%%%%%%%%%%%%%%%%%%%%%%%%%%%%%%%%%%%%%%%%%%%%%%%%%%%%%%%%%%%%%%%%%%%%
%\section{Introduction}
Achieving fine tuned control over rotation of intensity and azimuthal energy flow in structured light has emerged as a key requirement for enabling next generation optical micro-manipulation, quantum information transfer, and high-resolution imaging. Rotation in light fields arises from their intrinsic angular momentum, which governs how optical energy and momentum circulate and interact with matter. Fundamentally, the angular momentum of light comprises two distinct but interconnected components, spin angular momentum (SAM) and orbital angular momentum (OAM). SAM originates from the circular polarization of light and represents the intrinsic rotation of the electric field vector around its propagation axis, typically limited to values of $\pm\hbar$ per photon \cite{PhysRev.50.115, friese1998optical}, which can govern microscopic torque effects such as the rotation of birefringent particles but does not influence the macroscopic spatial structure of the beam. Consequently, SAM provides limited control over large-scale rotational dynamics or the directionality of energy flow in structured beams. In contrast, OAM arises from the helical phase structure of the optical field, expressed as $e^{i m \phi}$, where $m$ is the TC. This azimuthal phase dependence imparts a twist to the wavefront, creating a spiral Poynting vector that induces measurable rotation of intensity patterns and drives orbital motion in trapped particles. Unlike SAM, which is polarization-dependent and quantized only in two states, OAM can take any integer or fractional values of $m$, offering a scalable degree of freedom for controlling the rotational properties of light. However, while OAM provides a powerful handle on the macroscopic rotation of structured beams, it alone does not fully describe the underlying energy flow mechanisms responsible for observed rotational dynamics. The total rotational motion is determined by the transverse components of the Poynting vector, which combine contributions from both SAM and OAM as well as from their mutual coupling, particularly in tightly focused or non-paraxial beams \cite{BLIOKH20151}. In such regimes, spin–orbit interaction leads to conversion between SAM and OAM, altering local energy circulation and giving rise to complex internal rotations that cannot be predicted by TC alone. These limitations have motivated the exploration of beam configurations in which energy flow and rotation can be engineered independently of the OAM sign.

Among these, optical vortex beams, which carry a well-defined azimuthal phase singularity, serve as a fundamental platform for investigating the interplay between angular momentum and energy circulation. Each photon in such a beam carries an OAM of $m\hbar$, corresponding to a quantized twist in the phase front. The phase singularity at the beam center gives rise to a null intensity region, while the surrounding optical field exhibits an azimuthally varying phase gradient. The presence of this azimuthal phase gradient generates a transverse component of the Poynting vector, which forms a spiral energy flow around the propagation axis. The Poynting vector (S), represents the local energy flux density of the optical field. In vortex beams, the presence of the azimuthal phase term $m\phi$ introduces a tangential component $S_\phi$ in addition to the longitudinal component $S_z$. As a result, total energy flows no longer follow a straight path along the propagation direction but instead trace a helical trajectory that causes the optical energy and any trapped microscopic particle to experience a tangential momentum component, leading to measurable orbital rotation around the beam axis \cite{Bekshaev_2011}. Here, the ratio $S_\phi$/$S_z$ defines the helical pitch of the energy flow, a measure of how tightly the energy spirals as it propagates. The angular velocity of this motion is directly related to the local azimuthal phase gradient and therefore to the value of $m$. Consequently, beams with higher $\lvert m \rvert$ exhibit faster azimuthal energy circulation and stronger orbital torque on trapped particles, a property that forms the foundation for numerous applications in optical micromanipulation \cite{PhysRevLett.78.4713, padgett2011tweezers}, and angular momentum transfer \cite{friese1998optical, Bekshaev_2011, allen1992orbital, yao2011orbital}.

In both theoretical and experimental studies, a new class of discrete optical vortices (DOVs) have been recognized as an effective platform for engineering the rotational dynamics of structured light, alongside traditional continuous optical vortices (OVs). These DOVs have been observed in optical lattices of Bose–Einstein condensates \cite{fleischer2003observation}, 1D and 2D periodic photonic structures \cite{malomed2001discrete}, 1D ring networks of coupled lasers \cite{alexeyev2009linear}, coupled parametric oscillators \cite{wang2009formation}, and linear circular fibre arrays \cite{zhi2019comprehensive}. Certain applications, such as laser ablation, nonlinear frequency conversion, long-range optical communication, material processing, optical trapping and manipulation \cite{yao2011orbital, arita2013laser, PhysRevLett.67.3749}, require high power laser output which in a way can be achieved using more number of laser sites arranged in different concentric rings around a common center of origin of a DOV. Unlike their continuous counterparts, DOVs exhibit a discrete phase distribution across well-defined laser sites, producing a step-like phase variation around a central zero-intensity core. Building upon this concept, a phase-locked 1D ring array of lasers within a degenerate cavity has been shown to support the controlled generation of DOVs with arbitrary system size and TC \cite{Dev:21}. In the present study, we extend this principle to multiple concentric ring, enabling the generation of vortex beams with tunable rotational and spatial characteristics. Additionally, recent studies have highlighted that the introduction of asymmetry in the phase or amplitude of vortex beams leads to rotational behavior during propagation. For instance, asymmetric twisted vortex Gaussian Schell-model beams have been shown to exhibit crescent-shaped intensity profiles that rotate as they propagate, resulting from the coupling between the vortex phase and an additional twist phase within the correlation function \cite{hyde2024twisted}. Similarly, asymmetric Laguerre–Gaussian (LG) beams, generated by introducing a complex shift in the beam axis, demonstrate rotation of their intensity lobes \cite{singh2023tailoring}. However, these beams often suffer from spatial shape deformation upon propagation due to the absence of specific modal properties \cite{Kovalev:16}. These findings underscore that beam asymmetry inherently induces rotation, driven by a modified azimuthal phase gradient. However, achieving precise control over this rotation in terms of its direction, rate, and stability remains a key challenge.

Beam steering, i.e., the controlled rotation or redirection of a beam’s propagation direction at any desired transverse plane, fundamentally relies on the interaction of light (or electromagnetic waves) with materials or engineered structures that can impart spatially varying phase and amplitude responses. A wide variety of approaches have been demonstrated, such as spiral phase plates (SPPs), tunable liquid-based devices \cite{Cheng2021optical}, or metasurfaces built from subwavelength dielectric or plasmonic elements \cite{zhao2022meta}, and active phase-change materials such as vanadium dioxide (VO$_2$) allow electronically controlled beam steering \cite{hashemi2016electronically}. Across these platforms, the common principle is that, the beam rotation arises not from the mechanical movement, but from material responses, whether liquid reconfiguration, nanostructure-induced phase gradients, or electronic phase transitions, that establish the necessary wavefront gradient to steer the beam. However, traditionally, the direction of rotation of vortex beams in the generation plane has been regarded as an intrinsic property decided exclusively by the helicity (sign) of their TC \cite{allen1992orbital}. Addressing this, in the present work, we move beyond the conventional picture and demonstrate that, helicity is not the sole determinant of beam rotation in all the cases. Instead, a phase-engineered DOV framework based on concentric ring arrays, where the selective assignment of TCs across rings governs both handedness and the angular speed of rotation. This mechanism decouples rotational control from helicity, marking a shift toward purely phase-engineered beam steering without material mediation. Such a strategy provides a conceptually distinct pathway toward all-optical control of beam dynamics, highlighting the DOV system as a powerful alternative to material-assisted methods.

\section{Working principle of DOVs and generation of asymmetric vortices} \label{theo_desc}
\begin{figure*}[htbp!]
\centering
\includegraphics[width = 13 cm, keepaspectratio = true]{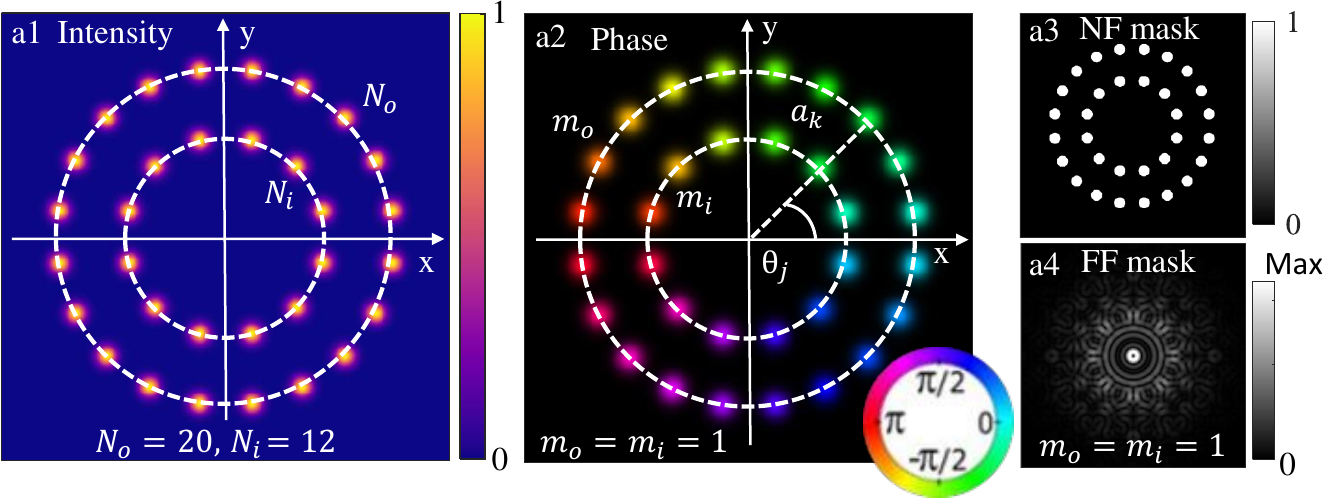}
\caption{Demonstartion of spatially multiplexed concentric discrete optical vortices. (a1--a2) Intensity and phase distributions of a DOV with concentric ring geometry. Here, $N_o$ and $N_i$ denote the number of lasers in the outer and inner rings, respectively, while $m_o$ and $m_i$ represent the corresponding TC (phase) configurations of each ring. (a3--a4) Near-field (NF) binary amplitude mask and far-field (FF) gray scale amplitude mask employed for generating a two-ring DOV with identical TC configurations across the concentric rings.}
\label{fig1}
\end{figure*}

\noindent
A DOV consists of a finite number of spatially distinct coherent optical sources, such as laser sites, beamlets, or waveguides, configured in a one-dimensional ring array. It features a central intensity null and exhibits stepwise phase circulation between adjacent sites, progressing either in a clockwise or counter-clockwise direction. A DOV with multiple concentric rings is demonstrated in Fig.~\ref{fig1}, consisting of 20 lasers in the outer ring and 12 lasers in the inner ring with specific initial phase distributions in the vortex configuration. In this figure, $N_o$ and $N_i$ denote the number of lasers in the outer and inner rings, respectively. Similarly, $m_o$ and $m_i$ represent the TCs (phase distributions) associated with the outer and inner rings of the DOV. These terminologies are used consistently throughout the remainder of this paper.

For the general case of DOV with concentric rings, the electric field distribution can be written as:
\begin{equation}
\begin{split}
E(x,y;z=0) = E_0
\sum_{k=1}^{M}
\sum_{j=1}^{N_k}
&\exp\!\left[
-\frac{(x-\alpha_{k,j})^2+(y-\beta_{k,j})^2}
{2\sigma_0^2}
\right] \\
&\times \exp\!\left(i\phi_{k,j}\right).
\end{split}
\label{eq1}
\end{equation}
where $(\alpha_{k,j}, \beta_{k,j}) = q_k (\cos\theta_j, \sin\theta_j)$ denote the radial positions of the $j^{\text{th}}$ laser beam in the $k^{\text{th}}$ ring. The radius of the $k^{\text{th}}$ ring is defined as 
\begin{equation}
q_k = \frac{d}{\sqrt{1 - \cos(2\pi/N_k)}}, \label{eq2}
\end{equation}
and the angular position and initial phase of the $j^{\text{th}}$ laser beam in the $k^{\text{th}}$ ring are given by 
\begin{equation}
\theta_j = \frac{\pi(2j-1)}{N_k}, 
\qquad
\phi_j = \frac{\pi m (2j-1)}{N_k}. \label{eq3}
\end{equation}
Here, $d$ denotes the centre-to-centre distance between two adjacent lasers in the same ring, $M = 1,2,\ldots$ defines the number of concentric rings,  and $N_k = 1,2,\ldots$ specifies the number of lasers in the $k^{\text{th}}$ ring. The TC of the beam in a specific ring is represented by $m$, where each laser ($j$) has the same amplitude $E_0$ and identical beam waist $\sigma_0$ but different initial phases $\phi_j$. 

To form a discrete optical vortex (DOV), the ring array must satisfy the periodic boundary condition $E_{j,k+N_k} = E_{j,k}$~\cite{pal2015phase}, meaning that as one completes a full rotation around the ring, the field at the final site matches that of the initial laser. 
In the case of a DOV, the phase difference between neighbouring lasers determines the total phase circulation around the ring, defining the topological charge as
\begin{equation}
\mathrm{TC} = \frac{1}{2\pi} \sum_{j=1}^{N_k} \arg\!\left(E_j^{*} E_{j+1}\right), \label{eq4}
\end{equation}
where $E_j$ represents the complex field amplitude of the $j^{\text{th}}$ laser site in the ring. It is important to note that, unlike continuous vortex beams, where TC in principle, can assume any integer value, a discrete ring supports only a finite set of allowed TCs, constrained by the number of laser sites such that  $|m| \leq N/2$ \cite{PhysRevA.83.063822}. This limitation originates from the discrete sampling of the azimuthal phase around the ring and constitutes a fundamental property of discrete vortex systems.

In this work, we adopted the phase-locked laser array scheme previously demonstrated in our earlier study \cite{Dev:21}, which utilizes a degenerate laser cavity with a 4f self-imaging configuration incorporating near-field (NF) and Fourier-field (FF) masks. This configuration enables independent lasing from multiple binary apertures arranged in ring geometry and enforces a stable, phase-locked state through spatial Fourier filtering at the far-field plane inside the cavity. By appropriately designing the NF and FF masks, the system facilitates the deterministic generation of DOVs with desired TC ($m$) and arbitrary system size ($N$). Leveraging this concept, we extend it to spatially multiplexed concentric ring arrays, enabling precise control over the rotational direction of intensity profile of the generated vortex beams. For numerical simulations, we employed a modified Fox–Li algorithm to achieve phase locked lasers within a degenerate cavity, enabling the generation of DOV with concentric ring configurations \cite{Tradonsky2016talbot}. Owing to the one round-trip propagation inside the degenerate cavity, the simulation involves the following operations:
\begin{equation}
E_{n+1}(x, y) = N F \!\left(\mathcal{F}^{-1}\left(FF\!\left(\mathcal{F}\!\left( G \times E_n(x, y) \right) \right) \right) \right),
\end{equation}
The process involves iterative propagation of the optical field within a degenerate cavity. Starting with the near-field distribution \(E_n(x, y)\), the field is first amplified by a spatially varying saturated gain \(G(x, y)\), which is then Fourier-transformed $(\mathcal{F}$) to the far-field domain, where a gray-scale amplitude mask (FF mask) modulates its spatial frequency components. An inverse Fourier transform ($\mathcal{F}^{-1}$) returns the field to the spatial domain, where a binary amplitude mask (NF mask) containing circular holes in a ring geometry applies to select desired regions. The updated field \(E_{n+1}(x, y)\) represents the state after one round trip, and the process repeats until it converges to a stable phase-locked state. Here, the saturated gain is described as
\begin{equation}
G(x, y) = \frac{G_0}{1 + \frac{I(x, y)}{I_{\text{sat}}}}, \label{eq6}
\end{equation}
where, \(G_0\) is the unsaturated gain, \(I_{\text{sat}}\) is the saturation intensity, and \(I (x, y) = E^2 (x, y)\) is the local intensity. In all our simulations, we have considered \(G_0 = 15\) and \(I_{\text{sat}} = 1000\). In this case, the simulation requires around $100$ iterations to converge to a desired steady-state of DOV.

The generation of DOVs with multiple concentric rings, each exhibiting identical TCs, is illustrated in Fig.\,\ref{fig1}. Figure\,\ref{fig1}(a1) shows the intensity distribution of the resulting DOV with \(N_o = 20\) and \(N_i = 12\). Figure\,\ref{fig1}(a2) display the phase distributions of DOVs with identical TCs (\(m_o, m_i) =(1, 1\)). Figure\,\ref{fig1}(a3) shows the near-field (NF) binary amplitude mask used to arrange the lasers in a two-ring concentric DOV configuration with the same system size as that of the intensity profile shown in Fig.\,\ref{fig1}(a1). Figure\,\ref{fig1}(a4) shows the far-field (FF) gray-scale amplitude mask used to find the respective phase distributions.

\section{Experimental Setup} \label{theo_exp}
\begin{figure*}[htbp]
 \includegraphics[width = 17.0 cm]{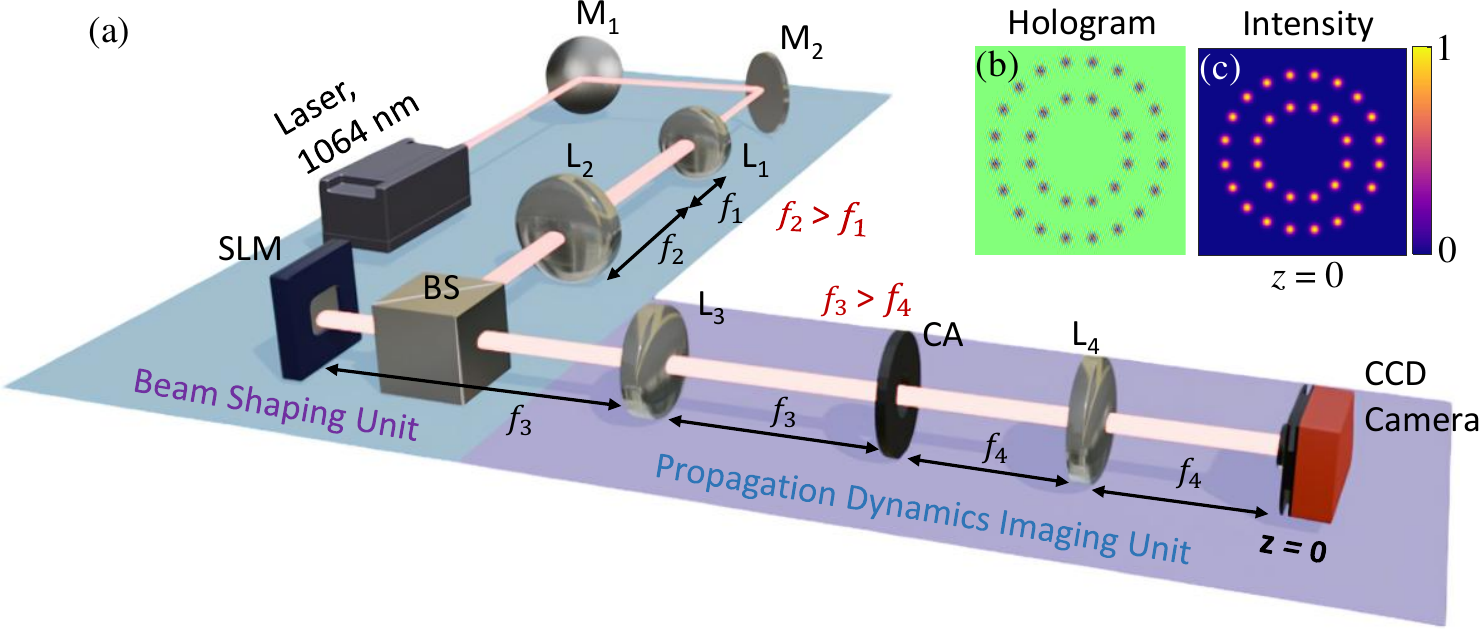}
\centering
\caption{Generation of spatially multiplexed concentric discrete optical vortices. (a) Schematic of the experimental setup used for the generation and characterization of concentric DOVs. A continuous-wave laser at 1064~nm is phase-modulated using a spatial light modulator (SLM) within the beam-shaping unit, followed by a beam splitter (BS), mirrors ($M_1$, $M_2$), and relay optics ($L_1$, $L_2$) forming a telescopic system ($f_2 > f_1$) to ensure proper illumination of the SLM. The structured beam is then directed to the propagation-dynamics imaging unit comprising lenses $L_3$ and $L_4$, a circular aperture (CA), and a lens assembly ($f_3 > f_4$) to probe free-space propagation. The resulting intensity evolution is recorded using a CCD camera at the observation plane ($z = 0$). (b) Phase hologram encoded on the SLM for generating concentric DOVs with the desired TC distribution, and (c) Experimentally generated intensity profile of the concentric DOVs at the initial plane.}
\label{fig2}
\end{figure*}
The modified Fox–Li algorithm mimics the experimental realization of a degenerate cavity laser, and has been tested and verified experimentally in several of the earlier works. However, for the experimental convenience to show the proof of concept, we have experimentally generated a DOV with two concentric rings from a computer-generated hologram (CGH) using a spatial light modulator (SLM). The schematic of the experimental setup is presented in Fig.\,\ref{fig2}(a).

%\begin{figure}[htbp]
%\centering
%\includegraphics[width=0.91\textwidth]{schematic _blender2.pdf}
%\caption{\textbf{Generation of spatially multiplexed concentric discrete optical vortices.} (a) Schematic of the experimental setup used for the generation and characterization of concentric DOVs. A continuous-wave laser at 1064~nm is phase-modulated using a spatial light modulator (SLM) within the beam-shaping unit, followed by a beam splitter (BS), mirrors ($M_1$, $M_2$), and relay optics ($L_1$, $L_2$) forming a telescopic system ($f_2 > f_1$) to ensure proper illumination of the SLM. The structured beam is then directed to the propagation-dynamics imaging unit comprising lenses $L_3$ and $L_4$, a circular aperture (CA), and a lens assembly ($f_3 > f_4$) to probe free-space propagation. The resulting intensity evolution is recorded using a CCD camera at the observation plane ($z = 0$). (b) Phase hologram encoded on the SLM for generating concentric DOVs with the desired TC distribution, and (c) Experimentally generated intensity profile of the concentric DOVs at the initial plane.}\label{fig2}
%\end{figure}
The experiments were performed using a phase-only spatial light modulator (SLM) with a resolution of $1920 \times 1080$ pixels and a pixel pitch of $8~\mu\mathrm{m}$. A collimated Gaussian laser beam at a wavelength of $\lambda = 1064~\mathrm{nm}$ was normally incident onto the SLM through a beam splitter (BS). The size of the input Gaussian beam was expanded six times by the magnifying telescopic combination of plano-convex lenses \(L_1~(f_1 = 5~\text{cm})\) and \(L_2~(f_2 = 30~\text{cm})\) to illuminate the entire screen of the SLM. On the SLM screen, a computer-generated phase hologram corresponding to the desired DOV is applied as shown in Fig.\,\ref{fig2}(b). The phase hologram displayed on the SLM modulates the amplitude and phase of the incident Gaussian laser beam, thereby splitting it into multiple laser beams arranged on two concentric one-dimensional (1D) ring arrays with prescribed phase distributions corresponding to specific vortex configurations. After reflection from the SLM, the modulated light is distributed among several diffraction orders. The desired DOV, carried by the first diffraction order, is spatially isolated using a circular aperture (CA) positioned at the Fourier plane of a telescope formed by plano-convex lenses \(L_3~(f_3 = 30~\text{cm})\) and \(L_4~(f_4 = 10~\text{cm})\). After spatial filtering, we obtain a desired DOV at the back focal plane of \(L_4\). Note that the beam size is reduced to one-third of its original value by the second telescopic configuration using \(L_3\) and \(L_4\) to ensure that the propagation characteristics of the generated DOV could be effectively captured within the active area of the used CCD camera. The experimental intensity distribution of generated spatially multiplexed concentric DOV is shown in Fig.\,\ref{fig2}(c). The concentric DOV architecture comprises two concentric one-dimensional (1-D) ring arrays containing $N_o = 20$ and $N_i = 12$ Gaussian laser emitters operating in the fundamental TEM$_{00}$ mode. A prescribed discrete phase distribution is assigned to the emitters in each ring to generate the desired vortex configuration. This is identical to a DOV obtained by phase-locking lasers in concentric 1-D ring arrays inside a degenerate cavity \cite{pal2015phase}.

For the concentric DOV configuration considered here, each Gaussian laser emitter operates with a beam waist of $\sigma_0 = 0.13~\mathrm{mm}$, and the center-to-center separation between adjacent emitters is chosen as $d = 0.56~\mathrm{mm}$ to facilitate experimental implementation.

\section{Results and discussions}
\label{results}
To demonstrate the formation of propagation invariant asymmetric modes and unconventional  rotational dynamics, we consider two spatially multiplexed concentric DOVs, each comprising a different number of constituent lasers and carrying either identical or distinct TCs on the inner and outer rings in various combinations. These different configurations give rise to several intriguing physical phenomena, including controlled clockwise or counter-clockwise rotation of transverse intensity distribution without altering the helicity (sign of TCs) of the DOVs, controlled splitting of higher-order TCs, and the formation of asymmetric vortex beams exhibiting propagation-invariant characteristics. 
%Furthermore, the rotational dynamics of these concentric DOVs exhibits striking similarities to those of magnetic skyrmions. 

\begin{figure*}[ht!]
 \includegraphics[width=12 cm, keepaspectratio=true]{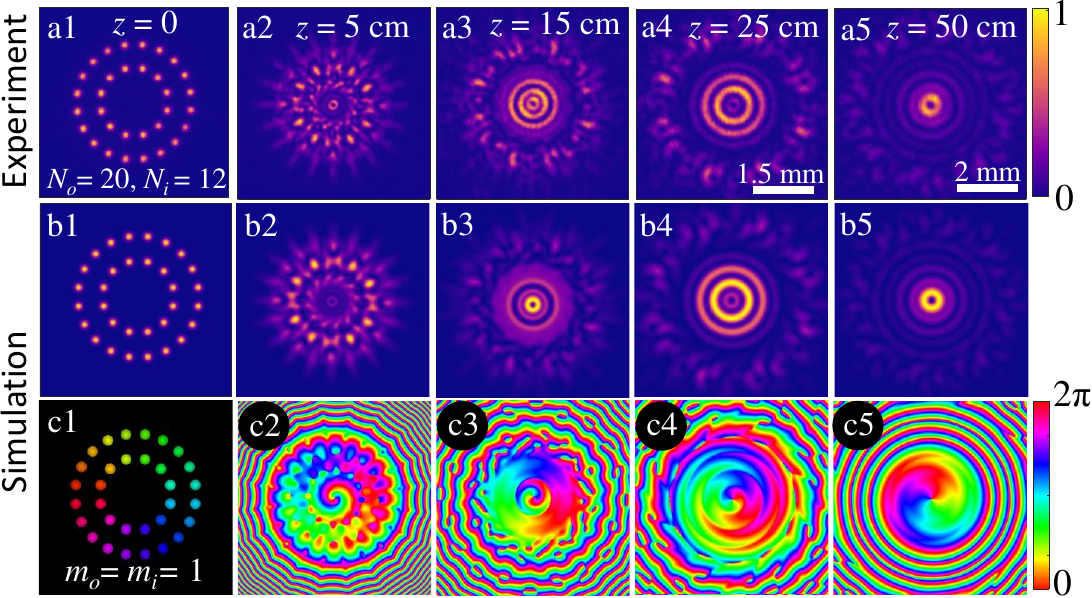}
\centering
\caption{Propagation dynamics of two concentric DOVs carrying identical TCs. For two concentric DOVs with $N_o = 20$ and $N_i = 12$, and TC $(m_o, m_i) = (1,1)$, experimental and simulated results of intensity distributions shown in (a1--a5), and (b1--b5) at different propagation distances ($z$), respectively. (c1--c5) Evolution of phase distribution at different propagation distances ($z$).}
\label{fig3}
\end{figure*}
\subsection{Formation of symmetric composite vortex beams}
When spatially multiplexed two concentric DOVs carrying identical TCs propagate, their fields progressively overlap, resulting in an interference-induced intensity redistribution. This interaction gives rise to a composite vortex beam characterized by concentric ring-shaped intensity profiles and a central dark core. The resulting intensity distribution remains symmetric in both the radial and azimuthal directions, while the phase retains a coherent helical structure. Figure\,\ref{fig3} presents the numerical and experimental results for two concentric DOVs with \(N_o = 20\) and \(N_i = 12\), carrying identical TCs (\(m_o, m_i) =(1, 1\)). Figures\,\ref{fig3}(a1--a5), and \ref{fig3}(b1--b5), show the experimental and simulated intensity distributions at different propagation distances ($z$), respectively. Figure\,\ref{fig3}(c1--c5) depicts the evolution of the phase distributions during propagation. As evident, the superposition of two concentric DOVs with identical TCs produces a structurally symmetric composite vortex beam with concentric ring-like intensity distributions and a well-defined helical phase profile. The concentric ring-like intensity pattern remains stable during propagation, indicating structural robustness arising from the shared TC on concentric DOVs. Furthermore, the experimental observations are in good qualitative agreement with the numerical simulations.

%\begin{figure}[ht!]
% \includegraphics[width=8 cm]{Fig_2.pdf}
%\centering
%\caption{Spectral evolution of a $500$\thinspace{fs} input pulse ($18$\thinspace{nJ}) at $1550$\thinspace{nm} through a $8$\thinspace{m} long GRIN-MMF with a core diameter of $50\thinspace\mu$m.}
%\label{fig2}
%\end{figure}

\subsection{Formation of asymmetric composite vortex beams}
\begin{figure*}[htbp]
 \includegraphics[width=16 cm]{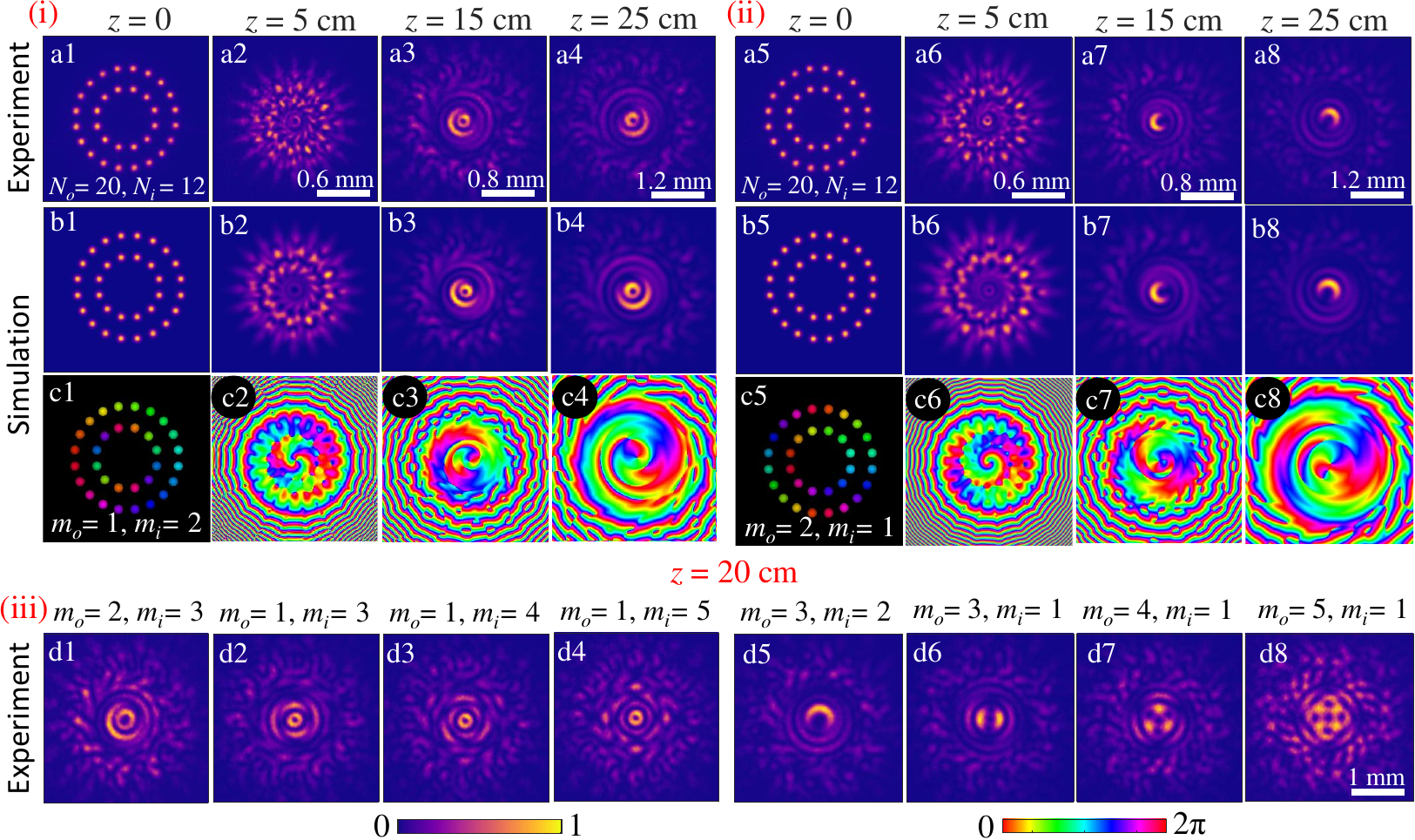}
\centering
\caption{Propagation dynamics of two-concentric DOVs carrying non-identical TCs. Experimental intensity evolution (first row), simulated intensity evolution (second row), and corresponding simulated phase evolution (third row) for concentric DOVs with $N_o = 20$ and $N_i = 12$, carrying TCs (i) (\(m_o,m_i)=(1,2)\) and (ii) (\(m_o,m_i)=(2,1)\). (iii) Experimental intensity distributions recorded at a propagation distance of $z=20\thinspace$cm for two concentric DOVs with different TC combinations: (d1) (\(m_o,m_i)=(2,3)\), (d2) (1,3), (d3) (1,4), (d4) (1,5), (d5) (3,2), (d6) (3,1), (d7) (4,1), (d8) (5,1).}
\label{fig4}
\end{figure*}
When spatially multiplexed concentric DOVs carrying different TCs propagate, the overlap of their optical fields leads to the formation of asymmetric composite vortex beams with non-uniform azimuthal intensity and phase distributions. This asymmetry further gives rise to a controlled rotation of the vortex structure about the propagation axis during free-space propagation. The experimental and numerical results are shown in Fig.\,\ref{fig4}. Figures\,\ref{fig4}(i) and \ref{fig4}(ii) illustrate the evolution of the intensity and phase distributions for two concentric DOVs with \(N_o = 20\) and \(N_i = 12\), carrying different combinations of TCs: (\(m_o,m_i)=(1,2)\) and (2,1), respectively. Figure\,\ref{fig4}(iii) presents the experimentally observed intensity distributions at $z=20\thinspace$cm for several TC combinations: \((m_o,m_i)=(2,3)\) [Fig.\,\ref{fig4}(d1)], (1,3) [Fig.\,\ref{fig4}(d2)], (1,4) [Fig.\,\ref{fig4}(d3)], (1,5) [Fig.\,\ref{fig4}(d4)], (3,2) [Fig.\,\ref{fig4}(d5)], (3,1) [Fig.\,\ref{fig4}(d6)], (4,1) [Fig.\,\ref{fig4}(d7)], and (5,1) [Fig.\,\ref{fig4}(d8)].

%\begin{figure*}[htbp]
% \includegraphics[width=16 cm]{Fig_4.pdf}
%\centering
%\caption{Propagation dynamics of two-concentric DOVs carrying non-identical TCs. Experimental intensity evolution (first row), simulated intensity evolution (second row), and corresponding simulated phase evolution (third row) for concentric DOVs with $N_o = 20$ and $N_i = 12$, carrying TCs (i) (\(m_o,m_i)=(1,2)\) and (ii) (\(m_o,m_i)=(2,1)\). (iii) Experimental intensity distributions recorded at a propagation distance of $z=20\thinspace$cm for two concentric DOVs with different TC combinations: (d1) (\(m_o,m_i)=(2,3)\), (d2) (1,3), (d3) (1,4), (d4) (1,5), (d5) (3,2), (d6) (3,1), (d7) (4,1), (d8) (5,1).}.
%\label{fig4}
%\end{figure*}

As evident, unequal TC distributions on the inner and outer rings of two-concentric DOVs lead to the formation of asymmetric composite vortex beams. Unlike conventional symmetric vortex beams, such as Laguerre–Gaussian (LG) modes, which are characterized by rotational symmetry, annular intensity profiles, and a single well-defined helical phase structure, the generated asymmetric vortices do not correspond to a unique eigenmode. Instead, they arise from the coherent superposition of concentric DOVs with dissimilar TCs, thereby breaking the rotational symmetry associated with standard vortex beams. A distinctive feature of these asymmetric vortices is the spatially non-uniform distribution of phase singularities and orbital angular momentum across the beam profile. Consequently, different regions of the beam exhibit different local phase gradients and effective TCs, giving rise to a richer and more intricate phase topology. Despite their modal non-uniqueness, these beams exhibit remarkable structural robustness. The asymmetric intensity pattern remains preserved during propagation, while the entire vortex structure undergoes a controlled rotation about the propagation axis. As shown in Figs.\,\ref{fig4}(i) and \ref{fig4}(ii), the beam maintains its overall shape and asymmetry over a considerable propagation distance, demonstrating propagation-invariant behavior accompanied by rotational dynamics. This combination of structural invariance and controllable rotation distinguishes the generated beams from conventional vortex modes and highlights their potential for applications requiring stable, shape-preserving optical fields, including off-axis optical manipulation, optical tweezing, particle transport, beam steering, and structured-light-based information encoding.

The proposed approach enables the systematic generation of a wide range of asymmetric optical vortices through the controlled assignment of distinct TCs to the concentric DOVs. Each DOV acts as an independent OAM channel, while the relative TC distribution governs the interference among the constituent fields, thereby enabling precise control over the resulting phase and intensity structures. By appropriately selecting the TCs of the inner and outer rings, vortex beams with tailored asymmetry and complex spatial profiles can be synthesized. The propagated intensity distributions shown in Fig.\,\ref{fig4}(iii) illustrate that even small variations in the TC combinations lead to markedly different asymmetric vortex patterns. These observations highlight the versatility of the proposed scheme and its ability to generate a broad range of customizable structured-light fields with controllable asymmetry.

\subsection{Emergence of azimuthal symmetry transition occurring in spatially multiplexed concentric DOVs}
\begin{figure*}[htbp]
 \includegraphics[width=12 cm]{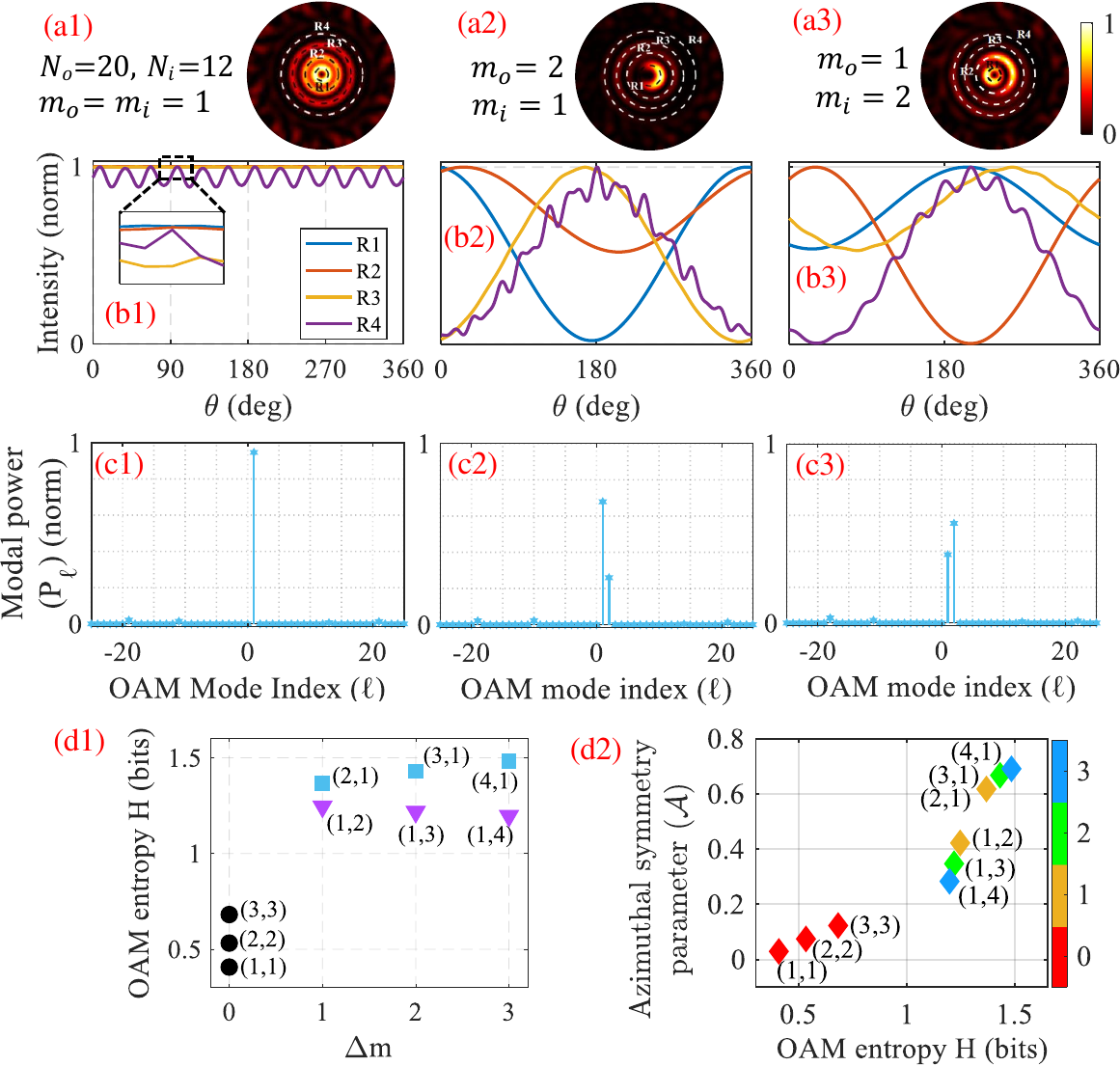}
\centering
\caption{Quantification of rotational symmetry breaking and OAM-mode diversification in concentric DOVs. (a1--a3) Propagated far-field intensity distributions of concentric DOVs with ($N_o=20$, $N_i=12$) number of lasers in the outer and inner rings, for the TC combinations $(m_o,m_i)=(1,1)$, $(2,1)$, and $(1,2)$, respectively. The white dashed circles (R$_1$--R$_4$) indicate the dominant radial intensity maxima selected for azimuthal intensity analysis. (b1--b3) Normalized azimuthal intensity distributions measured along the corresponding radial intensity rings.
(c1--c3) Normalized OAM power spectra ($P_\ell$), obtained from azimuthal Fourier decomposition of the propagated fields. While the equal-TC DOV is dominated by a single OAM mode, unequal-TC configurations exhibit redistribution of optical power among multiple OAM modes. (d1) OAM entropy ($H$), plotted as a function of the TC mismatch ($\Delta m=|m_o-m_i|$), for different TC combinations. (d2) Rotational symmetry-breaking parameter $S$ as a function of OAM entropy $H$, demonstrating the transition from low-entropy rotationally symmetric states to high-entropy symmetry-broken states, highlighting the role of OAM-mode diversification in the formation of asymmetric DOVs. The color bar represents the TC mismatch $\Delta m$.}
\label{fig5}
\end{figure*}

To explain the mechanism underlying the formation of symmetric and asymmetric composite vortex beams, we systematically analyzed the propagated beam structures generated from spatially multiplexed concentric DOVs with different combinations of TC assigned to the inner and outer rings. Representative intensity distributions for the configurations $(m_o, m_i)=(1,1)$, $(2,1)$, and $(1,2)$ are shown in Figs.\,\ref{fig5}(a1--a3). 
While the equal TC configuration preserves a nearly rotationally symmetric annular intensity profile, unequal TC configurations exhibit pronounced azimuthal intensity modulation, leading to the emergence of asymmetric vortex states.  To establish the generality of these observations, a comprehensive analysis was performed for a broad range of TC combinations. Whereas, the representative cases presented in Fig.\,\ref{fig5} were selected to illustrate the characteristic behavior of the symmetric and asymmetric beam families identified from this larger parameter space. It should be noted that the present analysis does not attempt to characterize the symmetry of the entire propagated optical field in a strict mathematical group-theoretic sense. Rather, the degree of symmetry considered here is defined in terms of the azimuthal intensity modulation of the physically relevant concentric ring structure, which constitutes the most physically significant feature of the propagated field. Importantly, the asymmetry does not manifest as a radial distortion of the beam envelope, rather it appears as a non-uniform redistribution of optical power along the azimuthal direction. To quantify this behavior, the dominant radial intensity maxima (R$_1$--R$_4$) were identified from the propagated intensity distributions and used as reference contours for azimuthal intensity analysis as illustrated in Figs.\,\ref{fig5}(a1--a3). The corresponding normalized azimuthal intensity cross-sections are shown in Figs.\,\ref{fig5}(b1--b3). For the equal-TC configuration, the intensity remains nearly invariant as a function of azimuthal angle for all dominant radial intensity rings, confirming the preservation of rotational symmetry. In contrast, unequal TC configurations exhibit substantial intensity modulation around the beam axis, indicating the breaking of rotational symmetry. For the concentric DOVs investigated here, the interference between the constituent discrete vortex rings governs the emergence of rotational symmetry breaking, while simultaneously redistributing the optical power among multiple OAM modes. To quantify this behavior, the propagated field was decomposed into its constituent OAM modes using the standard azimuthal Fourier decomposition \cite{Forbes:16}. The resulting normalized OAM power spectra $P_\ell$ are shown in Figs.\,\ref{fig5}(c1--c3). As evident, the equal-TC configuration is characterized by a dominant OAM family assigned to the concentric rings of DOV. However, for unequal TCs, the optical power becomes distributed among a broader set of OAM states. This redistribution originates from the interference between the phase structures imposed by the concentric rings and the discrete nature of the DOV.
The degree of modal diversification was quantified using the OAM Shannon entropy \cite{6773024, Volyar:22},
\begin{equation}
H = -\sum_{\ell} P_{\ell}\log_{2}(P_{\ell}),
\end{equation}
where $P_{\ell}$ denotes the normalized modal power of the OAM mode (See details in Appendix\,\ref{Appendix_A}). Physically, $H$ measures the uncertainty associated with the OAM spectra and therefore quantifies the effective number of OAM modes participating in the beam. A pure OAM state corresponds to $H=0$, whereas larger entropy values indicate increasing modal diversity. The dependence of the OAM entropy on the TC mismatch $\Delta m = |m_o-m_i|$, is presented in Fig.\,\ref{fig5}(d1). The equal-TC configurations consistently occupy a low-entropy regime, indicating that most of the optical power remains confined within a single OAM mode (eg., for $m_o = m_i=1$, the dominant OAM mode $\ell = 1$ retains $\sim93\%$ of the total modal power). In contrast, unequal-TC configurations exhibit substantially larger entropy values, demonstrating enhanced modal mixing. Interestingly, configurations possessing the same TC mismatch but opposite radial ordering, such as $(2,1)$ and $(1,2)$, yield noticeably different entropy values. This observation reveals that the beam dynamics are not governed solely by the magnitude of the TC mismatch. Rather, the radial location at which the angular momentum is introduced plays a crucial role in determining the resulting modal structure.
To quantify the degree of rotational symmetry directly from the intensity distributions, we define an azimuthal symmetry-breaking parameter $\mathcal{A}$ based on the modulation of the dominant intensity rings (R$_1$--R$_4$) shown in Figs.\,\ref{fig5}((a1)-(a3)). The higher values of $\mathcal{A}$ correspond to stronger departures from rotational symmetry, while the lower values indicate nearly uniform intensity around the beam axis (See details in Appendix\,\ref{Appendix_B}). The relationship between $\mathcal{A}$ and $H$ shown in Fig.\,\ref{fig5}(d2), reveal a clear separation between equal and unequal-TC configurations. Symmetric vortex states occupy a low-entropy, low-symmetry-breaking region, whereas asymmetric states populate a regime characterized by both enhanced modal diversity and stronger azimuthal intensity modulation. These results establish a direct link between OAM-mode diversification and the emergence of rotational symmetry breaking in concentric DOVs. Consequently, the radial allocation of TC provides an additional degree of freedom to tailor the modal content and spatial symmetry of discrete vortex beams.

\subsection{Unconventional rotational dynamics of spatially multiplexed concentric DOVs}

The rotational dynamics of both the phase and intensity profiles in vortex beams is fundamentally governed by its TC ($m$), which sets both the rotation rate and the direction of the transverse field evolution. This behavior is closely associated with the rotational Doppler effect \cite{10.3389/fphy.2022.938593}, where the frequency shift is expressed as $\Delta \omega = m \Omega$, with $\Omega$ denoting the angular rotation rate. In practical terms, when a vortex beam interacts with a rotating object, or when the beam itself is observed in a rotating reference frame, the phase evolution effectively experiences an angular rotation amplified by a factor of $m$. Consequently, the TC acts as a multiplicative gain, such that larger $|m|$ produces a steeper azimuthal phase gradient and a correspondingly stronger Doppler shift, while the sign of $m$ determines the sense of rotation (clockwise for $m>0$ and counter-clockwise for $m<0$) \cite{10.3389/fphy.2022.931131}. This amplification makes the rotational Doppler signature more pronounced and experimentally accessible, underscoring the central role of $m$ in governing rotational dynamics in structured light fields.

\begin{figure*}[htbp]
 \includegraphics[width = 12 cm]{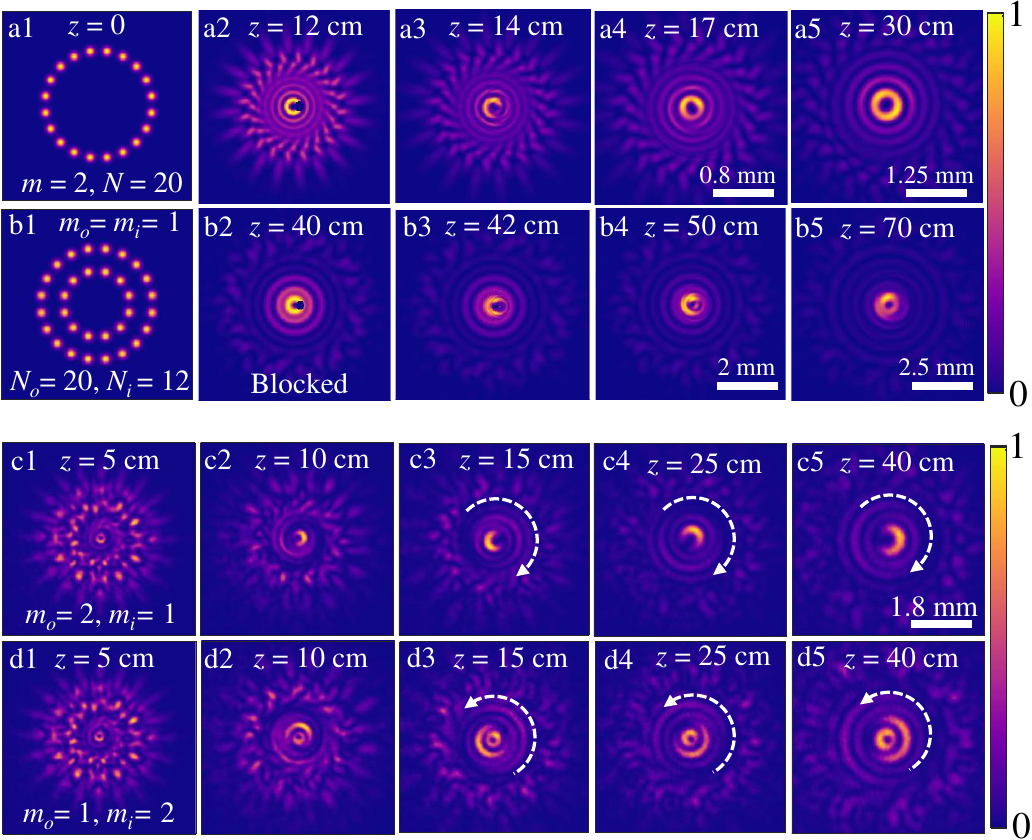}
\centering
\caption{Propagation-induced rotational dynamics of DOVs. (a1--a5) show the intensity evolution of a single ring DOV with TC $(m = 1)$ and number of lasers $(N=20)$, at different propagation distances after partial beam blocking at $z = 12$~cm and, panel (b1--b5) show the evolution of the beam intensity of a two-ring DOV with identical TCs $(m_o = m_i = 1)$ and number of lasers $(N_o = 20, N_i = 12)$ at different propagation distances after partial beam blocking at $z = 40$~cm, demonstrating the propagation-induced rotation of the symmetric vortex structure. 
(c1--c5) Intensity evolution of an asymmetric vortex beam exhibiting clockwise rotation for a concentric DOV with a higher TC in the outer ring $(m_o = 2, m_i = 1, N_o = 20, N_i = 12)$. 
(d1--d5) Reversal of the rotational direction to counter-clockwise obtained by interchanging the TC distribution between the rings $(m_o = 1, m_i = 2)$.}
\label{fig6}
\end{figure*}

To investigate this behavior in DOVs, we analyze the propagation dynamics of both single-ring and concentric-ring configurations with identical TCs across all rings. A small portion of the beam is intentionally blocked at $z = 12\,\mathrm{cm}$, and $z = 40\,\mathrm{cm}$ for the DOV of the single ring and two-rings, respectively, in order to visualize the rotational dynamics in symmetric vortex beams, as shown in Fig.\,\ref{fig6}(a1--a5) and Fig.\, \ref{fig6}(b1--b5), respectively. Upon further propagation, the truncated region progressively fills as the intensity distribution rotates, eventually reconstructing a complete ring. This representation effectively captures the intensity modulation and provides direct evidence of the rotational dynamics during propagation. These observations confirm that the symmetric DOV retains its vortex characteristics while undergoing continuous rotational evolution in free space.

Extending this concept to asymmetric DOVs, where concentric rings carry different TCs, introduces an additional degree of control. Since the TC directly governs the phase evolution and rotation rate, each ring in a multi-TC configuration behaves as an independent orbital angular momentum (OAM) channel. As a result, the rotational dynamics becomes tunable, and the rings may rotate at different angular velocities or even in opposite directions, depending solely on the sign and magnitude of their assigned TCs. Such counter-rotating behavior is consistent with previously reported multi-OAM beam interactions \cite{LiDanYuZhouZhangGaoLiXuYanYao+2023+4507+4517}, and highlights the versatility of concentric-ring DOVs for tailoring rotational field evolution.

A key observation emerging from our study is that the rotational dynamics of propagated beams are determined not only by the sign of TC \cite{bai2025dynamics}, but also by its spatial distribution across concentric rings. When the outer ring carries a higher TC than the inner ring (\(m_o = 2, \,m_i = 1, \,N_o = 20, \,N_i = 12\)), the generated asymmetric vortex beam rotates clockwise during propagation, as shown in Fig.\,\ref{fig6}(c1--c5). In contrast, reversing the same TC configuration, assigning the higher TC to the inner ring (\(m_o = 1, \,m_i = 2\)), produces an asymmetric vortex that rotates in the opposite (counter-clockwise) direction, as depicted in Fig.\,\ref{fig6}(d1--d5). Contrary to conventional expectations, our results reveal that the rotational sense of the resulting asymmetric vortex beam is governed not only by the absolute sign of the TCs, but also by the superposition, and their relative spatial hierarchy across the rings. This allows the rotation to reverse (clockwise or counter-clockwise) simply by redistributing the charges, marking a counterintuitive yet powerful degree of rotational control in discretized vortex systems.

\begin{figure*}[htbp]
 \includegraphics[width=12cm]{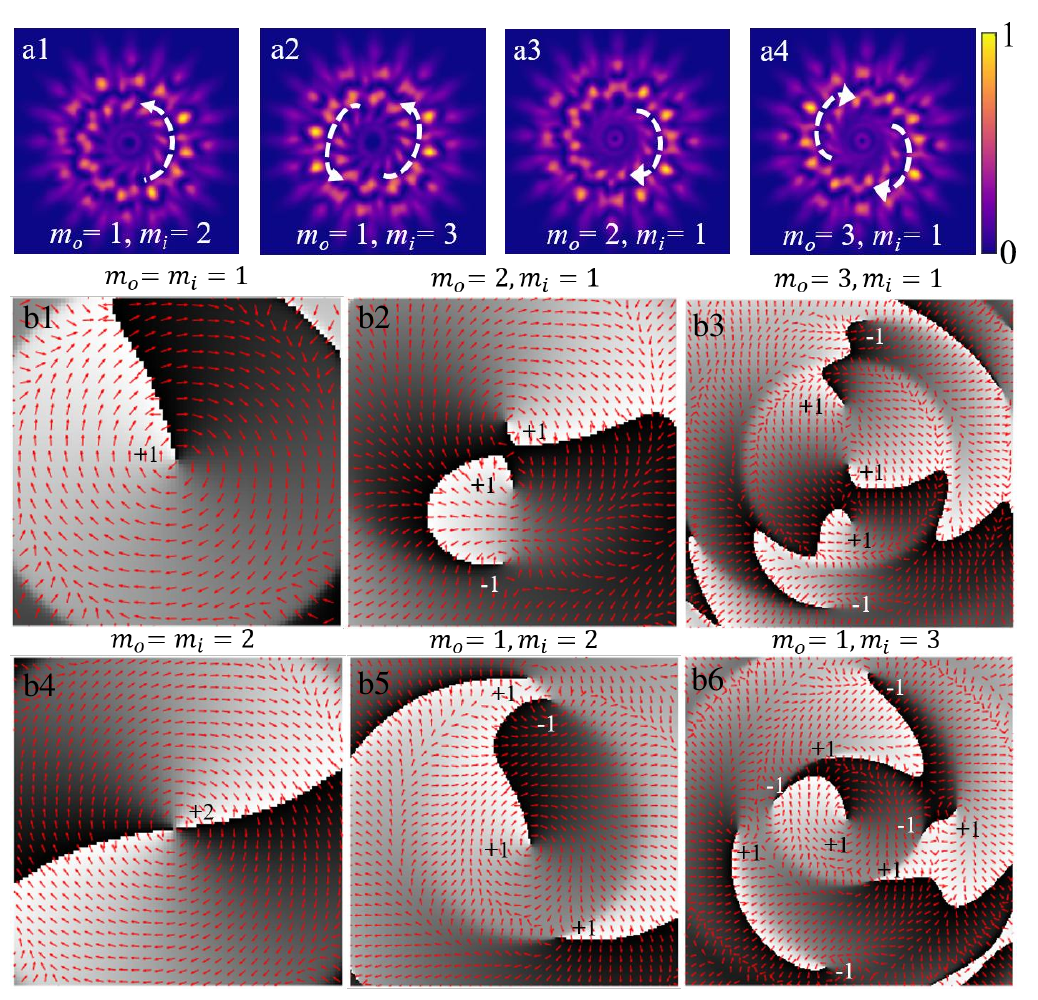}
\centering
\caption{(a1-a4) Two-concentric ring DOVs with varying TCs across the individual rings indicating clockwise and counter-clockwise intensity rotation illustrated using white dotted arrows. (b1-b6) Shows the direction of energy flow, indicated by red arrows, for a concentric DOV with different TCs across the rings.}
\label{fig7}
\end{figure*}

The magnitude contrast between the TCs assigned to each ring (e.g., \(m_o = 2, \,m_i = 1\), or vice-versa) introduces distinct phase velocities and azimuthal phase gradients across the radial extent of the beam, yielding a non-uniform evolution of phase and intensity upon propagation. In this regime, the net rotational behavior emerges from the radial distribution of angular momentum, where each ring acts as an independent OAM channel whose dynamical weight scales with both its charge and radial position. The outer ring, located at a larger radius, distributes its helical phase over a wider circumference, whereas the inner ring, though confined spatially, can dominate the angular momentum density when carrying a higher order TC. Again, when the inner ring possesses the higher TC (for equal signs), its accelerated azimuthal phase evolution imposes a radial phase gradient opposite to its local helicity, resulting in a global reversal of rotational sense during propagation, manifesting as counter-clockwise rotation of the intensity pattern (Fig.\,\ref{fig7}(a1-a2)). In contrast, when the higher TC resides in the outer ring, the radial and azimuthal phase gradients reinforce one another, producing clockwise rotation (Fig.\,\ref{fig7}(a3-a4)). Notably, reversing the signs of the same TC configuration preserves the spatial intensity morphology while inverting the rotation direction, demonstrating that the rotational handedness is ultimately encoded in the global phase helicity rather than in the intensity topology.

In optical vortex fields, a singular point organizes the entire field in the neighboring space. The exact Poynting vector represents the directional energy flux of an electromagnetic field and is defined as
\begin{equation}
\mathbf{S} = \frac{1}{\mu_0} \mathbf{E} \times \mathbf{B}
\label{eq:DOV}
\end{equation}

For simplicity, we consider a paraxial monochromatic coherent beam propagating along the $z$-axis, which can be expressed as
\begin{equation}
\mathbf{E}(\mathbf{r}) \approx E(x, z) e^{j k z}
\end{equation}
Where, the scalar complex amplitude of the paraxial optical field is denoted as 
\[E(x, z) = A(x) \exp[i\phi(x)],\]

which is a slowly varying function with respect to $z$. Note that in Eq.~\eqref{eq:DOV}, the time-dependent part $\exp[-j \omega t]$ (which has no effect on intensity) has been omitted, where $\omega$ is the angular frequency of the light wave. In quantum mechanics and optics, the probability or energy current is often written as 
\begin{equation}
\mathbf{S} \propto \operatorname{Im}(E^* \nabla E),
\end{equation}
which arises directly from the requirement of local conservation of probability (in quantum mechanics) or energy (in optics), formalized through the continuity equation. However, to obtain the energy flow (current or Poynting-like vector), the imaginary part is essential in most physical definitions, such as the probability current in quantum mechanics and paraxial optics.

Under the paraxial and scalar approximations typically employed in vortex beam optics, the time-averaged Poynting vector becomes parallel to the local wavevector. Hence, in the paraxial regime,, where the field predominantly propagates along a chosen direction (e.g., $z$), Maxwell’s equations allow a scalar field approximation for the optical beam \cite{d57e9d8bd5dc4619b6e512478ac542e3}. Since the local wavevector can be expressed as the phase gradient,
\[
\mathbf{k}_{\text{local}} = \nabla \phi(x, y, z),
\]
the transverse Poynting vector field is directly proportional to the product of the intensity and the gradient of the phase, given by
\begin{equation}
\mathbf{S}_{x,y} = \frac{1}{k} I(x, y) \nabla_{\!\perp} \phi(x, y),
\end{equation}

where $I(x, y)$ is the local beam intensity and $\phi(x, y)$ is the transverse phase gradient \cite{article50,Kumar:13,PhysRevLett.80.2586}.  

The transverse flow of energy is then visualized by evaluating the phase gradients $\nabla_{\!\perp} \phi(x, y)$, which serve as an approximation to the local direction of the Poynting vector. For clarity of presentation, these vectors are normalized to emphasize the direction of energy transport (perpendicular to the phase fronts); however, their density and scaling may optionally be weighted by intensity to indicate the relative strength of the flux. Using this relationship, we have demonstrated the direction of the energy flow of the asymmetric vortex beam for different combinations of TC in Fig.~\ref{fig7}(b1-b6).

Furthermore, for a regular, single-mode vortex beam (such as a pure LG-beam), the surrounding intensity profile does not physically rotate in a direction opposite to the core TC during free-space propagation. The handedness and structure of the intensity ring are fundamentally linked to the phase winding (TC) at the core, with no independent or contrary rotation possible in the absence of interference or nonlinear effects. Apparent cases where the intensity pattern seems to rotate in the opposite direction to the core TC can occur, but only under special circumstances, in engineered superpositions (such as interference of two beams with opposite or differing TCs), the resulting intensity features (petals or blades) can rotate in a direction opposite to the phase winding. This is a well-understood interference artefact and not a property of a single regular vortex beam. In nonlinear optical lattices or vortex ensembles, there can be collective effects where the central vortex (TC) and surrounding vortices are arranged to influence each other's rotational dynamics, sometimes allowing for stabilization or counteracting rotation within engineered structures. However, these scenarios involve more complex field configurations than a single regular vortex beam \cite{PhysRevLett.108.263903}.

%\begin{figure}[ht!]
 %\includegraphics[width=8 cm]{Fig_2.pdf}
 %\includegraphics[width=0.72\textwidth]{angular speed _sim _1.pdf}
%\centering
%\caption{Demonstrate exponential fit of the peripheral angular excursion $\Delta\theta$ as a function of outer TC $m_o$, with inner charge fixed at $m_i = 1$. The simulation data (blue circles) are well described by the function $\Delta\theta = A e^{-B m_o} + C$ (dotted red curve). The fit demonstrates that increasing outer TC suppresses angular excursion, consistent with stronger peripheral confinement and phase averaging. Spiral paths are embedded into the figure according to their respective TC pairs defined as $m_o m_i$.}
%\label{fig10}
%\end{figure}

\section{Conclusions}
\noindent
In summary, we have numerically synthesized the generation and propagation of spatially multiplexed concentric DOVs and experimentally verified their propagation behavior using a spatial light modulator (SLM), with strong correspondence observed between numerical predictions and measured results. By tailoring the spatial distribution of TCs across two concentric rings, we systematically explored the generation and propagation dynamics of both symmetric and asymmetric composite vortex beams. Configurations with identical TCs were found to evolve into rotationally symmetric vortex fields after certain free-space propagation, whereas non-identical TC assignments produced stable asymmetric vortex structures whose morphology depended sensitively on the specific charge pairing and  their radial allocation. A key outcome of this study is the discovery of a counterintuitive rotational response, contrary to the conventional expectation that the rotation direction of a vortex beam is determined exclusively by the sign of its TC, concentric DOVs enable controllable reversal of the net rotational direction without altering the sign of any charge. A higher TC in the inner ring can reverse the net angular momentum flow, leading to counter-clockwise rotation, whereas a higher TC in the outer ring drives clockwise rotation.  This behavior results from the interference between distinct azimuthal and radial phase gradients introduced by each ring, coupled with the non-uniform distribution of OAM across radial coordinates. The resulting differential OAM flow establishes an effective radial phase delay that dictates the global rotation direction of the composite vortex field, an effect that lies beyond the framework of conventional single-ring vortex theory. The demonstrated controllability of rotation in concentric DOVs, based on the superposition and coaxial propagation, provides a versatile mechanism for tailoring phase evolution in structured light and establishes a foundation for tunable topological control in next-generation photonic platforms without the risk of catastrophic collapse. 

%Recent studies have established that the rotational dynamics in skyrmion bags arises from the phase offset between the breathing oscillations of the inner and outer skyrmions, governed by the boundary-mediated interaction potential. A closely related mechanism emerges in concentric DOV architectures as well, where the angular speed of peripheral vortices is determined by the specific pairing of TCs distributed across radially separated rings. Furthermore, increasing the order of TC of the outer ring suppresses angular excursion by enhancing azimuthal phase confinement, an effect directly mirrors the suppressed rotational dynamics observed in skyrmion bags with elevated kink density in chiral magnetic systems. This cross-disciplinary correspondence reveals a unifying physical principle: rotational behavior in both optical and magnetic topological structures emerges from the coupled dynamics between inner and outer topological modes, modulated by the structural and phase complexity at the periphery. The demonstrated controllability of rotation in concentric DOVs, based on the superposition and coaxial propagation, provides a versatile mechanism for tailoring phase evolution in structured light and establishes a foundation for tunable topological control in next-generation photonic platforms without the risk of catastrophic collapse. 

\section*{Acknowledgements}
\noindent
We acknowledge financial support through the National Quantum Mission (NQM) of the Department of Science and Technology, Government of India. Aditya Narayana Jena acknowledges the fellowship support from IIT Ropar.

\appendix
\section{OAM decomposition and entropy analysis} \label{Appendix_A}

The emergence of asymmetric beam structures in concentric DOVs is accompanied by a redistribution of optical power among different OAM states. To quantify this redistribution, the propagated field is represented in cylindrical coordinates as
\begin{equation}
E(r,\theta) =
\sum_{\ell}
A_{\ell}(r)e^{i\ell\theta},
\label{eq:A1}
\end{equation}
where $\ell$ denotes the OAM mode index and $A_{\ell}(r)$ represents the radial amplitude associated with the corresponding OAM mode. The modal coefficients are obtained through azimuthal Fourier decomposition,
\begin{equation}
A_{\ell}(r) =
\frac{1}{2\pi}
\int_{0}^{2\pi}
E(r,\theta)e^{-i\ell\theta}d\theta.
\label{eq:A2}
\end{equation}
The modal power distribution is then defined as
\begin{equation}
P_{\ell} =
\frac{\int |A_{\ell}(r)|^{2}rdr}
{\sum_{\ell}\int |A_{\ell}(r)|^{2}rdr},
\label{eq:A3}
\end{equation}
such that
\begin{equation}
\sum_{\ell}P_{\ell}=1.
\end{equation}
The quantity $P_{\ell}$ therefore represents the fraction of total optical power carried by the OAM mode $\ell$.
To quantify the degree of modal diversification, we employ the Shannon entropy of the OAM spectrum,
\begin{equation}
H =
-\sum_{\ell}
P_{\ell}\log_{2}(P_{\ell}),
\label{eq:A4}
\end{equation}
where the logarithm is evaluated using base 2 and the entropy is therefore expressed in units of bits. The entropy measures the uncertainty associated with the OAM spectrum and serves as a measure of modal diversity. A pure OAM state yields $H=0$, while broader modal distributions correspond to larger entropy values.
Physically, the OAM entropy characterizes the extent to which optical power is distributed among multiple OAM families. Consequently, it provides a quantitative measure of the complexity of the underlying angular momentum structure and serves as a useful indicator of the transition from ordered symmetric vortex states to more complex asymmetric beam configurations.

\section{Quantification of azimuthal symmetry breaking parameter (S)} \label{Appendix_B}
The symmetric and asymmetric beam states reported in this work are distinguished through their azimuthal intensity distributions. Since concentric DOVs inherently contain multiple radial intensity rings, rotational symmetry must be evaluated independently at each radial location.
For a given radius $r_i$, the azimuthal intensity profile is defined as
\begin{equation}
I(r_i,\theta).
\label{eq:C1}
\end{equation}
For a perfectly rotationally symmetric beam,
\begin{equation}
I(r_i,\theta) =
\mathrm{constant},
\qquad
\forall \theta.
\label{eq:C2}
\end{equation}
Any azimuthal variation therefore represents a deviation from rotational symmetry.
To quantify this variation, the modulation index associated with the $i$th dominant intensity ring is defined as
\begin{equation}
M_i =
\frac{\sigma_{\theta}(I)}
{\langle I\rangle_{\theta}},
\label{eq:C3}
\end{equation}
where
\begin{equation}
\langle I\rangle_{\theta} =
\frac{1}{2\pi}
\int_{0}^{2\pi}
I(r_i,\theta)d\theta
\label{eq:C4}
\end{equation}
is the azimuthal mean intensity and
\begin{equation}
\sigma_{\theta}(I) =
\sqrt{
\frac{1}{2\pi}
\int_{0}^{2\pi}
\left[
I(r_i,\theta) -
\langle I\rangle_{\theta}
\right]^2
d\theta
}
\label{eq:C5}
\end{equation}
is the corresponding standard deviation.
The modulation index therefore measures the relative intensity fluctuation around the beam axis. A perfectly rotationally symmetric ring yields $M_i=0$, while larger values indicate stronger azimuthal modulation.
Since different radial rings contribute unequally to the observed beam structure, a global symmetry-breaking parameter is obtained through an intensity-weighted average,
\begin{equation}
\mathcal{A} =
\sum_i
w_iM_i,
\label{eq:C6}
\end{equation}
where
\begin{equation}
w_i =
\frac{P_i}
{\sum_i P_i}
\label{eq:C7}
\end{equation}
and $P_i$ denotes the peak intensity associated with the $i$th radial ring.
The weighting procedure suppresses the influence of weak diffraction-induced outer rings while emphasizing the dominant structural features responsible for the visible beam morphology. Consequently, the parameter $\mathcal{A}$ provides a robust measure of azimuthal symmetry breaking that is directly linked to the experimentally observable intensity distribution.

%%%%%%%%%%%%%%%%%%%%%%%%%%%%%%%%%%%%%%%%%%%%%%%%%%%%%%%%%%%%%%%%%%%%%%%%%%%%%%%%%%%%%%%
\bibliography{bib_file}{}
\bibliographystyle{apsrev4-1}
\end{document}